\documentclass[12pt]{article}
\usepackage{epsf}
\usepackage{amsmath}
\usepackage{graphics}
\usepackage{cite}

\renewcommand{\thefootnote}{\#\arabic{footnote}}
\begin{document}

\newcommand{\gtrsim}{ \mathop{}_{\textstyle \sim}^{\textstyle >} }
\newcommand{\lesssim}{ \mathop{}_{\textstyle \sim}^{\textstyle <} }

\newcommand{\rem}[1]{{\bf #1}}

\renewcommand{\thefootnote}{\fnsymbol{footnote}}
\setcounter{footnote}{0}
\begin{titlepage}

\def\thefootnote{\fnsymbol{footnote}}

\hfill August 2016\\
\vskip .5in
\bigskip
\bigskip

\begin{center}
{\Large \bf Angular Momentum of Dark Matter Black Holes}

\vskip .45in

{\bf Paul H. Frampton\footnote{email: paul.h.frampton@gmail.com\\
homepage:www.paulframpton.org}}

{15 Summerheights, 29 Water Eaton Road, Oxford OX2 7PG, UK.}

\end{center}

\vskip .4in
\begin{abstract}
\noindent
The putative black holes which may constitute all the dark matter
are described by a Kerr
metric with only two parameters, mass M and angular momentum J. There
has been little discussion of J since it plays no role in the upcoming
attempt at detection by microlensing. Nevertheless J does
play a central role in understanding the previous lack of detection, especially 
of CMB distortion.
We explain why bounds previously derived from lack of CMB distortion
are too strong for primordial black holes
with J non-vanishing.
Almost none of the dark matter black holes can be
from stellar collapse, and nearly all are primordial,
to avoid excessive CMB distortion.
\end{abstract}

\end{titlepage}

\renewcommand{\thepage}{\arabic{page}}
\setcounter{page}{1}
\renewcommand{\thefootnote}{\#\arabic{footnote}}

\newpage

\section{Introduction}

\bigskip

\noindent
The Milky Way galaxy in which we reside lies within a large approximately
spherical halo of dark matter (DM) which does not experience the strong or
electromagnetic interactions, nor as we shall assume here the weak interactions.
The popular idea that the dark matter constituent is a WIMP with weak 
interactions was born out of supersymmetry which lacks any support
from extensive LHC data on pp scattering which probed the energy
regime where signs of SUSY were most expected. Die-hard SUSY theorists
may still have hope, but it is not premature to entertain the assumption
that the WIMP does not exist.

\bigskip

\noindent
With no WIMP one is led to astrophysical MACHOs and then confronted
with the constraint from BBN that no more than 20\% of the DM can be baryonic.
This means that to make 100\% of the DM we cannot use compact objects
such as white dwarfs, neutron stars, brown dwarfs and unassociated planets.
Nor is it possible to use black holes which are the result of gravitational collapse
of baryonic stars.

\bigskip

\noindent
There is, however, a second type of black hole which is
formed primordially (PBH) during the radiation era.
To form 100\% of the DM we must therefore use PBHs. Since the resultant black
holes of the two types are indistinguishable, can we use, say,  20\% of the gravitational
collapse variety and 80\% of PBHs? The surprising answer is no. One result of
the present paper is that if there is such a mixture of black holes the vast majority, 
well over 99\%, must be PBHs and only a tiny fraction can be the result of
gravitational collapse. This strong result comes from a study of X-rays and
associated CMB distortion.

\bigskip

\noindent
Focusing on the Milky Way halo where we can most easily detect the PBHs, we
already know from earlier searches, especially \cite{MACHO},
that MACHOs with masses $M \leq 20M_{\odot}$ can make
up no more that 10\% of the halo dark matter.  At the high mass end,
we know from \cite{Xu} that MACHOs with $M \geq 10^5 M_{\odot}$
endanger the disk stability. For the Milky Way halo therefore one is led
to consider intermediate mass (IM) black holes PIMBHs in the mass range

\bigskip

\begin{equation}
20 M_{\odot} \leq M_{PIMBH} \leq 10^5 M_{\odot}
\label{PIMBHmass}
\end{equation}
\bigskip
 
\noindent
for the DM constituents. This leads to a {\it plum pudding} model \cite{FramptonDM}
for the
Milky Way halo, named after Thomson's atomic model \cite{Thomson}, where
for the DM halo the plums are PIMBHs with masses satisfying Eq.(\ref{PIMBHmass}) and
the pudding in this case is rarefied gas, dust and just a few luminous stars.

\bigskip

\noindent
The formation of PBHs with masses as large as Eq.(\ref{PIMBHmass}) and much larger
is known to be mathematically possible during the radiation era. An existence theorem
is provided by hybrid inflationary models \cite{FramptonInflation}. One specific prediction of hybrid
inflation is a sharply-peaked PBH mass function. If we need a specific PIMBH mass, 
we shall use a calligraphic ${\cal PIMBH}$ defined by  $M_{{\cal PIMBH}} \equiv 100 M_{\odot}$
exactly. This is merely an example and extension to the whole range of Eq.(\ref{PIMBHmass})
can also be discussed.

\bigskip

\noindent
The cosmic time $t_{PBH}$ at which a PBH is formed has been estimated \cite{CKSY} to be

\begin{equation}
t_{PBH} \simeq \left( \frac{M_{PBH}}{10^5M_{\odot}} \right) ~~ seconds
\label{tPBH}
\end{equation}

\noindent
so that the PIMBHs in Eq.(\ref{PIMBHmass}) are formed in the time window
$0.0002s \leq t_{PIMBH} \leq 1.0s$
with the special case $t_{{\cal PIMBH}} \simeq 0.001 s$. In terms of red shift ($Z$),
this corresponds to
\begin{equation}
5 \times 10^{11} \geq Z_{PIMBH} \geq 5 \times 10^9
\label{ZPIMBH}
\end{equation}
with the special case $Z_{{\cal PIMBH}} \simeq 2 \times 10^{11}$. 
The formation of BHs which are not primordial, which we shall
denote without an initial $P$ or ${\cal P}$, necessarily occurs {\it after} star formation which conservatively
occurs certainly only for very different redshifts satisfying 
\begin{equation}
Z_{BH} \leq 1400
\label{ZBH}
\end{equation}
The sharp difference in the red-shifts of Eq.(\ref{ZPIMBH}) and Eq.(\ref{ZBH}) will become 
important when we discuss the reasons for previous non-detection, the angular momentum of PIMBHs and BHs,
and the central issue of possible CMB distortion by X-rays.

\bigskip

\noindent
As already mentioned, by using the mathematical models in \cite{FramptonInflation}, it is possible to form PBHs 
not only in the PIMBH mass range of Eq. (\ref{PIMBHmass}) but also Primordial Super Massive
Black Hole (PSMBHs) in the mass range
\begin{equation}
10^5 M_{\odot} \leq M_{PSMBH} \leq 10^{17} M_{\odot}
\label{PSMBHmass}
\end{equation}
where the upper limit derives from the formation time $t_{PSMBH}$ given by Eq. (\ref{tPBH}) staying
within the radiation-dominated era. We shall discuss the higher mass range Eq( \ref{PSMBHmass}) later in the paper.

\bigskip

\noindent
Finally for this Introduction, we recall that in a microlensing experiment, {\it e.g.} using the LMC or SMC for 
convenient sources, microlensing by halo PIMBHs, and assuming a typical transit velocity $200km.s^{-1}$, the time duration
of the microlensing light curve can be estimated \cite{Pacz} to be approximately
\begin{equation}
\tau \simeq \left( \frac{M_{PIMBH}}{25 M_{\odot}} \right)^{\frac{1}{2}}  ~~ years
\label{duration}
\end{equation}
which we note is close to one year and two years, respectively, for lens masses 
$25M_{\odot}$ and $100 M_{\odot}$. For reference, the highest duration such light curve
detected by the MACHO Collaboration which published in the year 2000 \cite{MACHO}
corresponded to $M_{PIMBH} \simeq 20 M_{\odot}$.

\newpage

\section{Kerr Metric and Period $\tau$}

\noindent
The PIMBHs are described by a Kerr metric \cite{Kerr} which has the form in 
Boyer-Lindquist $(t, r, \theta, \phi)$ coordinates, after defining
$\alpha = \frac{J}{M}$, $\rho^2 = r^2 + \alpha^2 \cos^2 \theta$ and $\Delta = r^2-2Mr +\alpha^2$,

\begin{eqnarray}
ds^2 &=& - \left( 1 - \frac{2Mr}{\rho^2} \right) dt^2 - \left( \frac{4Mr\alpha \sin^2\theta}{\rho^2} \right) d\phi dt + \left( \frac{\rho^2}{\Delta} \right) dr^2 \nonumber \\
& & + \rho^2 d\theta^2 +\left( r^2 + \alpha^2 + \frac{2Mr \alpha^2 \sin^2 \theta}{\rho^2} \right) \sin^2 \theta d\phi^2
\label{Kerr}
\end{eqnarray}

\bigskip

\noindent
In Eq.(\ref{Kerr}), there are two free parameters, $M$ and $J$. By reputation, analytic calculations
building on Eq. (\ref{Kerr}) can be impossibly difficult, usually leading to numerical techniques.

\bigskip

\noindent
In this paper, we shall need only order-of-magnitude estimates for the rotational period $\tau$ and, in the
next Section, for the angular momentum $J$. These will suffice to make our point about concomitant
X-ray emission. The solution is axially symmetric and the radius at the pole $\theta = \frac{\pi}{2}$
is the same as the Schwarzschild radius $R = 2M$. For other values of $\theta$ the black hole radius
is smaller than the static one and the rest of the static would-be sphere is filled out by an ergosphere whose equatorial
radius is also $R=2M$.  

\bigskip

\noindent
To proceed with our estimate we shall take the equatorial velocity of the ergosphere to have magnitude $V = 0.1c$
and use Newtonian mechanics to estimate the rotation period $\tau$ as simply

\begin{equation}
\tau = \left( \frac{2\pi R}{V} \right)
\label{period}
\end{equation}

\bigskip

\noindent
For the Sun, we have $2M_{\odot} \simeq 3 ~ km$ so that for a black hole of mass 
$M = \eta M_{\odot}$ and therefore radius $R \simeq 3 \eta ~ km$ Eq.(\ref{period}) is, for
$V=0.1c = 3 \times 10^4 km.s^{-1}$,

\begin{equation}
\tau = \left( 2 \times 10^{-4} \pi \eta \right) ~~ seconds
\label{tau}
\end{equation}

\noindent
Some values of $\tau$, estimated by this method, are shown in the third column of 
our Table.

\bigskip

\begin{table}[htdp]
\begin{center}
\begin{tabular}{|c|c|c|c|}
\hline
Astrophysical & Mass  & Period $\tau$   & Angular Momentum ${\cal J}$ \\
object       &  solar masses  & seconds  & $kg.km^2.s^{-1}$    \\
\hline
\hline
Earth & $M_{\oplus} = 6 \times 10^{24} kg$          &  24 hours                       &   $1.1 \times 10^{27}$        \\
\hline
Sun &   $M_{\odot} = 2 \times 10^{30} kg$      &    25 days           &  $1.1 \times 10^{36}$       \\
\hline
\hline
PIMBH & $20 M_{\odot}$ & $0.013 s $ & $3.0 \times 10^{37}$  \\
\hline
${\cal PIMBH}$ &  {\cal $100 M_{\odot}$} & {\cal $0.063 s$} &  $7.2 \times 10^{38}$  \\
\hline
PIMBH & $1000 M_{\odot}$  & $0.63s$ &  $7.2 \times 10^{40}$ \\
\hline
PIMBH & $10^4 M_{\odot}$  & $6.3s$ &  $7.2 \times 10^{42}$ \\
\hline
PIMBH & $10^5 M_{\odot}$ & $63s$ & $7.2 \times 10^{44}$ \\
\hline
\hline
PSMBH (M87) & $6 \times 10^9 M_{\odot}$ & $3.8 \times 10^6 s$ & $2.6 \times 10^{54}$  \\
\hline
\hline
\end{tabular}
\end{center}
\label{tauJ}
\end{table}

\section{Angular Momentum ${\cal J}$}

\bigskip

\noindent
Let us define the dimensionless angular momentun ${\cal J} \equiv J / kg.km^2.sec^{-1}$. We
are interested in order of magnitude estimates of ${\cal J}$ for the PIMBHs and PSMBHs. The value of ${\cal J}$ for astrophysical objects is necessarily
a large number so to set the scene we shall estimate ${\cal J}$ for the Earth ${\cal J}_{\oplus}$
and for the Sun ${\cal J}_{\odot}$.

\bigskip

\noindent
The parameters for the Earth are radius $R_{\oplus} \simeq 6300 km$, period $\tau_{\oplus} \simeq 86400 s$, mass
$M_{\oplus} \simeq 6 \times 10^{24} kg$,
hence angular velocity $\omega_{\oplus} = 2\pi /\tau_{\oplus}$ and moment of inertia $I_{\oplus} = \frac{2}{5} M_{\oplus}R_{\oplus}^2$
so an estimate is ${\cal J}_{\oplus} \sim I_{\oplus} \omega _{\oplus} \simeq 1.1 \times 10^{27}$.  For the Sun the similar calculation
using $R_{\odot} \simeq 700,000 km$, $\tau_{\odot} \simeq 25 days$, $M_{\odot} \simeq 2 \times 10^{30} kg$ gives ${\cal J}_{\odot} \simeq 1.1 \times 10^{36}$.

\bigskip

\noindent
For the black holes, the value of ${\cal J}$ is proportional to $\eta^2$ where $\eta = (M / M_{\odot})$. A similar estimate
to that for the Earth and Sun gives ${\cal J} \simeq 7.2 \times 10^{34} \eta^2$, which provides the remaining entries in our Table.

\section{CMB Distortion}

\noindent
Because of rotational invariance, angular momentum is conserved. The ${\cal J}$
of a compact astrophysical object will not change dramatically unless there is
an extremely unlikely event like a major collision. For example, the Earth and the
Sun in the first two rows of our Table were formed $4.6$ billion years ago. Their 
respective  angular momenta ${\cal J}_{\oplus}$ and ${\cal J}_{\odot}$ have
remained essentially constant all of that time. According to Eq.(\ref{tPBH}),
the PIMBHs listed in the next five rows of our Table were all formed at time
$t \leq 1s$ and their angular momenta have therefore remained roughly constant for
the last $13.8$ billion years since then.

\bigskip

\noindent
In detecting the dark matter, let us focus on the special case ${\cal PIMBH}$
with $M=100M_{\odot}$. The ${\cal PIMBH}$ was formed, accordng to
Eq.(\ref{tPBH}), at time $t=10^{-3} s$ and rotates with period $t\simeq 63ms$,
thus rotating $\sim 16$ times per second and with an absolute angular momentum
$\sim 6 \times 10^{11}$ times that of the Earth and $\sim 600$ times that of the Sun.
There is no known reason that ${\cal J}_{{\cal PIMBH}}$ would change
significantly after its formation.

\bigskip

\noindent
These remarks about angular momentum
are salient to resolving the contradiction between
the dark matter proposal in \cite{FramptonDM} and the limits on halo MACHOs
derived earlier by Ricotti, Ostriker and Mack (ROM) in \cite{ROM} on 
the basis of X-ray emission and CMB distortion.

\bigskip

\noindent
In \cite{FramptonDM} the proposal was made that the Milky Way dark halo
is a plum pudding with, as ``plums", PIMBHs in the mass range of Eq.(\ref{PIMBHmass})
making up $100\%$ of the dark matter. On the other hand,
in Figure 9 of ROM \cite{ROM}, there is displayed an upper limit of
less than $0.01\%$ of the dark matter for this mass range
of MACHO. Thus, it would seem that at least one of \cite{FramptonDM}
and \cite{ROM} must be incorrect? The conclusion of the present paper is
that ROM \cite{ROM} is correct for stellar-collapse black holes
but is not applicable to the model of \cite{FramptonDM} which
employs primordial black holes. This issue
was discussed also in \cite{ACDF}.

\bigskip

\noindent
This ROM upper limit arises from the lack of any observed departure
of the CMB spectrum from the predicted black-body curve or of any
CMB anisotropy. ROM calculated the accretion of matter on to the
MACHOs, the emission of X-rays by the accreted matter and then the
downgrading of these X-rays to microwaves by cosmic expansion
and more importantly by Compton scattering from electrons.

\bigskip

\noindent
A crucial assumption made by ROM \cite{ROM} is that the accretion on to
the MACHO can be modeled as if the MACHO has zero angular momentum
$ J = 0$. The justification for this assumption is based on 
earlier work by Loeb \cite{Loeb} who studied the collapse of gas clouds
at redshifts $200 \leq Z \leq 1400$. Such collapse can form compact
objects, eventually black holes, but during the collapse angular
momentum is damped out from the electrons by Compton scattering
with the CMB.

\bigskip

\noindent
From Loeb's discussion, the resultant black holes will have $J=0$
and this appears to underly why ROM \cite{ROM} used the Bondi-Hoyle
model\cite{BHmodel} which presumes spherical symmetry for accretion.
This is justified for stellar-collapse black holes by the arguments
of Loeb\cite{Loeb} and therefore the upper bounds derived by ROM
are applicable.

\bigskip

\noindent
There is evidence that the Bondi-Hoyle model of
accretion is not, by contrast, applicable to spinning PSMBHs, in particular the
one at the centre of the large galaxy M87. In \cite{M87A,M87B}
Bondi-Hoyle \cite{BHmodel} is used to calculate the number
of X-rays expected from the accreted material near M87.
In the case of M87 the X-rays
are experimentally measured. The conclusion is striking: that the
measured X-rays are less by several orders of magnitude than
predicted by Bondi-Hoyle theory.

\bigskip

\noindent
This supports the idea that the SMBHs such as that in M87
are primordial, so we list PSMBH(M87) in the final row of 
our Table. The ROM
constraints apply to black holes which originate from gravity 
collapse of baryonic stars. Collecting this fact, together with the
ROM limit of $\leq 10^{-4}$ of the dark matter for MACHOs, implies that
$99.99\%$ of the dark matter black holes are primordial, formed
during the radiation era.

\newpage 

\section{Discussion}

\bigskip

\noindent
The plum pudding model for the dark halo proposed in \cite{FramptonDM}
arose from a confluence of theoretical threads including study of the entropy
of the universe and the knowledge of how to form PBHs with 
many solar masses as in Eqs. (\ref{PIMBHmass}) and (\ref{PSMBHmass}). Nevertheless it was
the weakening of the argument for WIMPs which was most
decisive,

\bigskip

\noindent
The strongest objection to the MACHOs in \cite{FramptonDM} has been based on
the X-rays and the CMB distortion 
as calculated by ROM \cite{ROM}. In the present paper we have
attempted to lay this criticism to rest by noting that ROM assumed $J=0$
and that the putative PIMBHs have not only many times the Solar mass
but also many times the Solar angular momentum. This appears to us to render
the ROM constraints inapplicable to the PIMBHs. On the other hand, they
do apply to stellar-collapse black holes which implies that almost none
($\leq 0.01\%$) of the dark matter black holes are of that type.
To decide whether dark matter really is PIMBHs will require their
detection by a dedicated microlensing experiment.

\bigskip

\noindent
Examples
of PSMBHs may already have been observed in galactic cores and
quasars. Other PSMBHs can play the role of dark matter in clusters
and may well be detectable by other future lensing experiments. 
There is
also the upper mass range contained in Eq.(\ref{PSMBHmass}). 
Although masses of PSMBHs up to a few times
$10^{10} M_{\odot}$ may have already been observed in quasars, there
are what could be called Primordial Ultra Massive Black Holes (PUMBHs)
with masses between $10^{11}$ and $10^{17}$ solar masses which
might exist within the visible universe.

\bigskip
\bigskip

\noindent
\section*{Acknowledgement}
We thank Bernard Carr and George Chapline for useful discussions.

\end{document}